\documentclass[aps,prl,twocolumn,preprintnumbers,10pt,showpacs,nofootinbib]{revtex4-1}

\usepackage[utf8]{inputenc}
\usepackage{amsmath,amsfonts,epsfig,color,latexsym}
\usepackage{amssymb}
\usepackage[english]{babel}
\usepackage{epsfig,psfrag}
\usepackage{slashed}
\usepackage[normalem]{ulem}
\usepackage{xcolor}

\usepackage{booktabs}
\usepackage{array,multirow,graphicx}

\usepackage{soul,xcolor}
\usepackage{hyperref}
\setstcolor{red}
\usepackage{comment}

  \usepackage{graphicx}
  \usepackage{axodraw}
  \usepackage{subfigure}
  \usepackage{placeins}
  \usepackage{natbib}

  \usepackage{soul}

\newcommand{\be}{\begin{equation}}
\newcommand{\ee}{\end{equation}}

\newcommand{\bea}{\begin{eqnarray}}
\newcommand{\eea}{\end{eqnarray}}
\newcommand{\bfig}{\begin{figure}}
\newcommand{\efig}{\end{figure}}
\newcommand{\bc}{\begin{center}}
\newcommand{\ec}{\end{center}}

\usepackage{enumitem}
\setlist[itemize]{leftmargin=*}

\graphicspath{ {fig/} }

\allowdisplaybreaks[4]

\renewcommand{\thefootnote}{\fnsymbol{footnote}}

\begin{document}

\title{NLO Corrections to Light-Quark Mixed QCD-EW Contributions to Higgs Production}

\author{Matteo Becchetti}
\email{matteo.becchetti@unito.it}
\affiliation{Dipartimento di Fisica, Universit\`a di Torino and INFN Sezione di Torino, Via Pietro Giuria 1, I-10125 Torino, Italy}

\author{Roberto Bonciani}
\email{roberto.bonciani@roma1.infn.it}
\affiliation{Dipartimento di Fisica, Universit\`a di Roma ``La Sapienza'' and 
INFN Sezione di Roma, 00185 Roma, Italy}

\author{Vittorio Del Duca}
\email{delducav@itp.phys.ethz.ch}
\affiliation{ETH Z\"{u}rich, Institut f\"{u}r theoretische Physik, Wolfgang-Pauli
str. 27, 8093, Z\"{u}rich, Switzerland}

\author{Valentin Hirschi}
\email{valentin.hirschi@gmail.com}
\affiliation{ETH Z\"{u}rich, Institut f\"{u}r theoretische Physik, Wolfgang-Pauli
str. 27, 8093, Z\"{u}rich, Switzerland}

\author{Francesco Moriello}
\email{fmoriell@itp.phys.ethz.ch}
\affiliation{ETH Z\"{u}rich, Institut f\"{u}r theoretische Physik, Wolfgang-Pauli
str. 27, 8093, Z\"{u}rich, Switzerland}

\author{Armin Schweitzer}
\email{armin.schweitzer@phys.ethz.ch}
\affiliation{ETH Z\"{u}rich, Institut f\"{u}r theoretische Physik, Wolfgang-Pauli
str. 27, 8093, Z\"{u}rich, Switzerland}

\begin{abstract}
 We present for the first time the exact NLO QCD corrections to the light-quark part of the mixed QCD-EW contributions to Higgs production via gluon fusion at LHC13, with exact EW-boson mass dependence.
The relevant two-loop real-emission matrix element is computed using a dynamic one-dimensional series expansion strategy whose stability and speed allows for a numerical phase-space integration using local IR subtraction counterterms.
For $\mu_R=\mu_F=M_H$, we find:
\begin{flalign}
\sigma^{(\alpha_s^2\alpha^2+\alpha_s^3\alpha^2)}_{g g\rightarrow H+X} &=
    1.467(2)^{\;+18.7\%}_{\;-14.6\%}\;(\mu_R\;\text{var.})\;\pm 2\%\;(\text{PDF}) \ \mathrm{pb} \,,\nonumber
\end{flalign}
which we use to provide the best result including an estimate of suppressed contributions:
\begin{align}
    \sigma^{(\text{EW},\textrm{best})}_{p p\rightarrow H+X} = 2.11 &\pm 0.28 \  (\textrm{theory}) \ 
    \mathrm{pb}.\nonumber
\end{align}

\end{abstract}

\maketitle

After the discovery of the Higgs boson with the Large Hadron Collider (LHC) at CERN in 2012~\cite{Aad:2012tfa,Chatrchyan:2012ufa}, the work of the LHC community has focused on the study of the Higgs sector, which provides a stringent test of the Standard Model of particles (SM) and a fertile environment for the search of New Physics (NP) signals 
\cite{Harlander:2013oja,Banfi:2013yoa,Azatov:2013xha,Grojean:2013nya,Schlaffer:2014osa,Buschmann:2014twa,Dawson:2014ora,Buschmann:2014sia,Ghosh:2014wxa,Dawson:2015gka,Langenegger:2015lra,Azatov:2016xik,Grazzini:2016paz}.

A key ingredient for predictions of Higgs observables is the accurate knowledge of the Higgs production cross section in gluon fusion, which at the LHC is by far
the dominant production mode.
The coupling of the Higgs boson to gluons is mediated by a heavy-quark loop.
The Higgs production cross section was computed at
leading order in the '70s~\cite{Georgi:1977gs},
and at next-to-leading-order (NLO) in strong coupling constant $\alpha_s$
in the '90s~\cite{Graudenz:1992pv,Spira:1995rr}.
NLO QCD corrections are sizable ($\sim 80 - 100\%$) undermining
the reliability of the perturbative expansion in $\alpha_s$ of the production cross section.
The next-to-next-to-leading-order (NNLO)~\cite{Harlander:2002wh,Anastasiou:2002yz,Ravindran:2003um} and the next-next-to-next-to-leading-order (${\rm N^3LO}$)~\cite{
Anastasiou:2015ema,Mistlberger:2018etf} corrections in $\alpha_s$ have been computed in the Higgs Effective Field Theory (HEFT) approach, i.e. in the limit of a top quark much heavier than the Higgs boson, $M_T\gg M_H$, with all other quarks taken as massless, which replaces the loop-mediated coupling by an effective tree-level coupling.
The NNLO corrections turned out to be significant ($\sim 10 - 20\%$) but with a reduced scale-dependent uncertainty.
The ${\rm N^3LO}$ corrections turn out to be small ($\sim 4 - 6\%$)~\cite{Anastasiou:2016cez}, with a renormalization/factorization scale variation of less than 2\%.

The high accuracy of the ${\rm N^3LO}$ corrections calls for the evaluation of finite quark-mass effects and electroweak contributions. Finite quark-mass effects are known through NLO~\cite{Spira:1995rr,Harlander:2005rq,Anastasiou:2006hc,Aglietti:2006tp,Bonciani:2007ex,Anastasiou:2009kn,Anastasiou:2020qzk} and contribute a $\sim -7\%$ change~\cite{Anastasiou:2009kn} to
the cross section. 
Although the relevant ingredients - double-virtual~\cite{Harlander:2019ioe,Czakon:2020vql,Prausa:2020psw}, 
real-virtual~\cite{Bonciani:2016qxi,Bonciani:2019jyb,Frellesvig:2019byn} 
and double-real~\cite{DelDuca:2001fn,Budge:2020oyl} - of the computation of the finite quark-mass effects at NNLO are available, such a computation has not been performed yet.
At NNLO, top-quark mass effects have been estimated through a power expansion in $M_H/M_T$~\cite{Harlander:2009mq,Pak:2009dg,Harlander:2012hf} and found to be $\sim 1\%$. Light-quark mass effects, in particular the top-bottom interference, are not yet known at NNLO.

Gluon induced Electroweak (EW) effects arise at two loops, i.e. at ${\mathcal O}(\alpha^2 \alpha_s^2)$. They are due to the gluons coupling to EW bosons $V = W, Z$ through a quark loop, followed by the gauge coupling of the EW bosons to the Higgs boson. Mixed QCD-EW contributions were calculated for the light-quark loop~\cite{Aglietti:2004nj,Aglietti:2004ki,Degrassi:2004mx}, for the heavy-quark loop~\cite{Actis:2008ug} and with full quark-mass dependence \cite{Actis:2008ug}, and found to increase the ${\rm N^3LO}$ cross section by about 2\%  \cite{Anastasiou:2016cez}. Since this increase is of the order of the residual QCD uncertainty, it is important to compute the NLO corrections in $\alpha_s$. 
Because the largest part ($\sim 98\%$~\cite{Degrassi:2004mx})
of the increase at ${\mathcal O}(\alpha^2 \alpha_s^2)$ is due to the light-quark part of the mixed QCD-EW contributions, the evaluation of the NLO corrections has been aimed at the light-quark part. 
These corrections were evaluated in the limit that the Higgs mass is much smaller than the EW boson masses, $M_H\ll M_V$~\cite{Anastasiou:2008tj}, and turned out to be sizable.

The ${\mathcal O}(\alpha^2 \alpha_s^3)$ corrections consist of three parts: the one-loop $2 \to 3$, the three-loop $2 \to 1$, and the two-loop $2 \to 2$.
In \cite{Hirschi:2019fkz},
the one-loop $2 \to 3$ processes were computed and found to yield a negligible contribution.
The three-loop contribution was evaluated analytically and expressed in terms of multiple polylogarithms (MPLs)~\cite{Bonetti:2016brm}. In \cite{Bonetti:2018ukf}, the soft part of the two-loop $2 \to 2$ process was added. In \cite{Anastasiou:2018adr}, the total cross section was evaluated in the small EW-boson mass limit, $M_V\ll M_H$. 
 The planar MIs for the two-loop $gg \to Hg$ process with the exact EW-boson mass were published in \cite{Becchetti:2018xsk}
and recently, in \cite{Bonetti:2020hqh}, the complete helicity amplitudes, including the non-planar diagrams, were presented. The calculation was done analytically, expressing the results in terms of MPLs. In this work we perform an independent computation of the MIs by using the series solution method of \cite{Francesco:2019yqt}.

In this letter, we present the NLO QCD corrections to the total cross section for Higgs production via gluon fusion at the LHC, due to the light-quark part of the mixed QCD-EW contributions, with exact EW boson mass dependence.

\section{Calculation}
\paragraph{Loop amplitudes}
 The computation of the two-loop $gg \to Hg$ amplitude $\mathcal{M}_{gg\rightarrow Hg}^{(\alpha_s^3\alpha^2)}$ is performed by using the series expansion method of \cite{Francesco:2019yqt}. Specifically, we reduce the amplitude to master integrals (MIs) by using computer programs \cite{Maierhoefer:2017hyi, Smirnov:2014hma,Lee:2013mka} for the solution of Integration-By-Parts identities (IBPs) \cite{Chetyrkin:1981qh, Tkachov:1981wb, Laporta:2001dd} and Lorentz-Invariance identities (LI) \cite{Gehrmann:1999as}. By taking advantage of these identities we define a system of differential equations \cite{Kotikov:1990kg, Remiddi:1997ny, Gehrmann:2002zr, Argeri:2007up} for a basis of canonical MIs \cite{Henn:2013pwa}. 
 The canonical basis is found by using the methods of \cite{Henn:2013pwa, Henn:2014qga, Argeri:2014qva, Lee:2014ioa, Georgoudis:2016wff, Gehrmann:2014bfa, Becchetti:2017abb} and the corresponding differential equations are solved in terms of generalised power series as described in \cite{Francesco:2019yqt}. The system of differential equations uniquely defines the solution when imposing boundary conditions at a special kinematic point. We consider the infinite EW-boson mass limit as our initial boundary point. The numerical evaluation of the relevant scattering amplitudes by means of the generalised power series approach is well suited for Monte-Carlo phase space integrations. Moreover, the analytic continuation of the generalised power series to the physical regions is fully algorithmic.

The series solution strategy can be summarised as follows. We transport the integrals from a known boundary point to a phase-space point of interest, by solving the differential equations in terms of generalised power series along the line connecting the pair of points. This is done dynamically for every new phase-space point. When the line crosses a physical threshold, the analytic continuation is defined by assigning a vanishing imaginary part to the line parameter, in accordance to Feynman prescription. By construction, the non-analytic terms of the series are logarithms and rational powers of the line parameter, and their analytic continuation is elementary. In order to improve the efficiency of the evaluations, we consider a pre-computed grid of about 5K physical phase-space boundary points. In this way the series solution can be found along lines connecting pairs of points separated by a relatively short distance. In general, these lines cross fewer singular points of the differential equations, and they require, for fixed precision, lower truncation order of the series, considerably reducing the average evaluation time. In this work we consider truncated power series which guarantee a precision for the numerical evaluations of the integral basis of at least $16$ digits after the decimal point. The average time for one form factor evaluation on one CPU core is $\mathcal{O}(1\textrm{min})$, ranging from $\mathcal{O}(30\textrm{sec})$ up to $\mathcal{O}(10\textrm{min})$ for input kinematic configurations featuring large scale hierarchies.

We verified our computation against \cite{Bonetti:2020hqh} and report benchmark values for the relevant matrix elements in the appendix. As for the three- and two-loop form factors of the  $gg \to H$ process ($\mathcal{M}_{gg\rightarrow H}^{(\alpha_s^3\alpha^2)}$ and $\mathcal{M}_{gg\rightarrow H}^{(\alpha_s^2\alpha^2)}$ respectively) we use the results presented in~\cite{Bonetti:2017ovy,Bonetti:2016brm}.

\paragraph{Phase-space integration and infrared regularisation}
Infrared (IR) divergences are locally subtracted using two different paradigms for cross-validation: first, the FKS subtraction scheme~\cite{Frederix:2009yq} and second, a modified version~\cite{Lionetti:2018gko,DelDuca:2019ctm} of the {\sc\small CoLoRFul}~\cite{Somogyi:2006cz,Somogyi:2009ri} scheme at NLO. To the best of our knowledge, this is the first time that a two-loop matrix element with implicit phase-space IR divergences has been numerically integrated using local subtraction counterterms.\\
To achieve this, we encoded the loop amplitudes discussed in the previous paragraph as form factors of effective vertices in a {\sc\small UFO}~\cite{Degrande:2011ua} model.
We then created a custom plugin made publicly available\footnote{\url{https://bitbucket.org/aschweitzer/mg5_higgs_ew_plugin/} or\\ \url{http://madgraph.physics.illinois.edu/Downloads/PLUGIN/higgsew.tar.gz}} for the {\sc\small MadGraph5\_aMC@NLO} programme~\cite{Alwall:2014hca} allowing for the generation of a standalone library for the evaluation of all matrix elements entering our computation. 
We finally customised the generation output of {\sc\small MadGraph5\_aMC@NLO} for the NLO QCD correction of inclusive Higgs hadroproduction within the Higgs Effective Theory (HEFT) in order to accommodate the aforementioned matrix elements as well as an offline parallelisation pipeline.

\begin{figure}[ht!]
\centering
\includegraphics[width=7.cm]{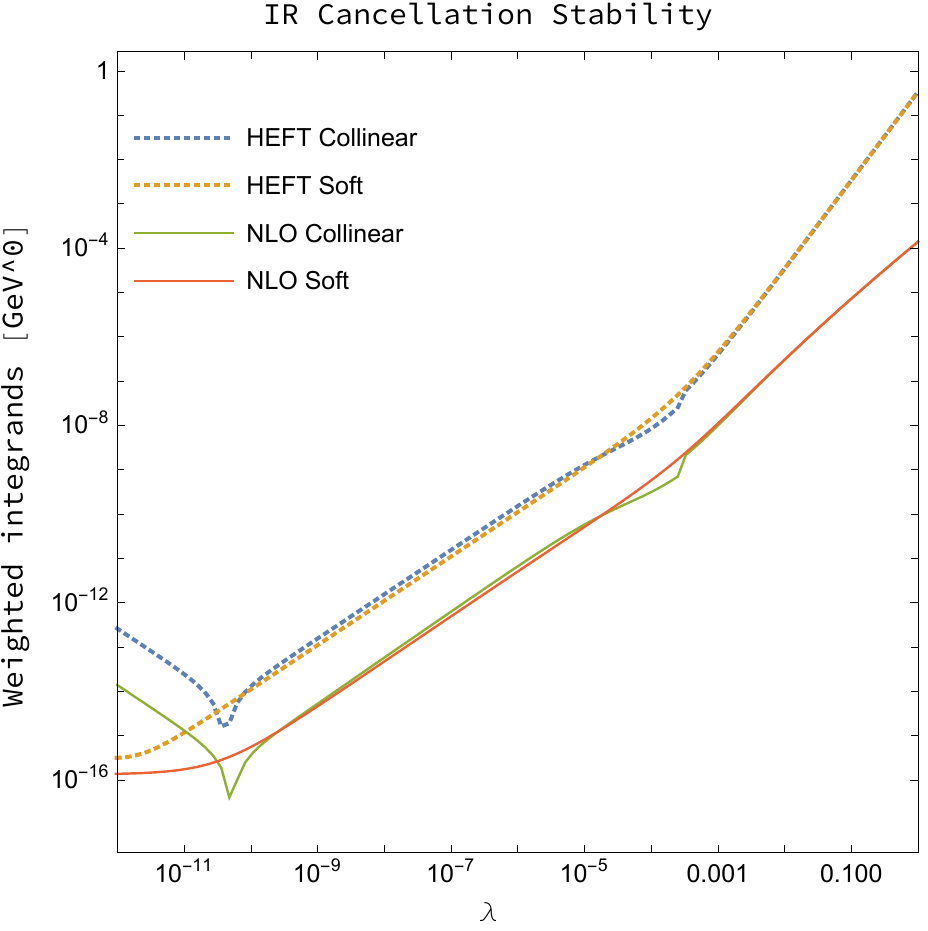}
\caption{\label{fig:IR_stability} Numerical stability of the 2-loop real-emission matrix element, locally subtracted with our modified implementation of {\sc\small CoLoRFuL}, compared to their HEFT tree-level counterpart when approaching the soft and collinear limits. The approach parameter $\lambda$ is defined so that the scaling of the real-emission matrix when approaching the IR limit is $\lambda^{-1}$. The weighted integrand shown includes the Jacobian of the parameterisation so that it must scale like $\lambda^{\alpha}$ with $\alpha>\frac{1}{2}$ in order to be integrable.
}
\end{figure}

We validated our results against those obtained using a modified version of the NLO implementation~\cite{Somogyi:2009ri} of the {\sc\small CoLoRFul} subtraction scheme rendered suitable for the computation of an Higgs inclusive cross-section and implemented in the private extension of {\sc\small MadGraph5\_aMC@NLO} already featured in ref.~\cite{Hirschi:2019fkz}.
In that variant, the local soft counterterm uses a mapping recoiling against initial states (see ref.~\cite{DelDuca:2019ctm} and sect. 5.3.3 of ref.~\cite{Lionetti:2018gko}) and \emph{locally} identically cancels against its soft-collinear counterpart. This is a consequence of the fact that our real-emission matrix element only features soft emission from an initial-initial dipole.
The initial-final collinear counterterms alone are thus sufficient to regularise the infrared (IR) divergences involved in our computation. We showcase the stability of the real-emission matrix element in fig.~\ref{fig:IR_stability} by investigating the quality of its cancellation against local IR counterterms both in the collinear and soft limit. We find stability on par with that obtained in HEFT when considering tree-level real-emission matrix elements, where the limiting factor is the double-precision accuracy of the input kinematics.
We also tested the independence of our result on the arbitrary cutoff of the local IR collinear counterterms (parameter $y_0^\prime$ in ref.~\cite{Somogyi:2009ri}). 
\\
We verified that our two independent implementations of the {\sc\small FKS} and {\sc\small CoLoRFuL} IR subtraction procedure give consistent results.

\section{Results}
We carry out our computation in the Standard Model (SM) using the relevant input parameters given in table~\ref{tableParams}.
\renewcommand*{\arraystretch}{1.2}
\begin{table}[h!]
\begin{center}
\begin{tabular}{ll|ll}
Parameter & value & Parameter & value
\\
\hline 
PDF set & {\tt PDF4LHC15\_nlo\_30}\;\; & $\mu_R=\mu_F$ & $M_H$
\\
$\alpha_{S}(\mu_R)$ & As per PDF set. & $G_F$ & $\frac{\pi \alpha}{\sqrt{2}M_W^2(1-M_W^2/M_Z^2)}$
\\
$\sqrt{\hat{s}}$ & \tt{13} TeV & $\alpha^{-1}$ & \tt{132.507}
\\
$M_Z$ & \tt{91.1876} & $V^{CKM}_{ij}$ & $\delta_{ij}$
\\
$M_{W}$ & \tt{80.3845} & $M_H$ & \tt{125.09}
\end{tabular}
\end{center}
\caption{\label{tableParams} Standard Model parameters used for obtaining all numerical results presented in this letter. Dimensionful parameters are given in GeV unless indicated otherwise. All particle widths are set to zero.}
\end{table}

We remind the reader that we consider for the W-exchange only the first two light-quark generations, whereas the Z-exchange receives also contributions from massless $b$-quarks as discussed in \cite{Bonetti:2016brm,Bonetti:2017ovy}. We furthermore neglect the contribution from $g g \rightarrow H q \bar{q}$ (computed to be below $0.1$ pb in ref.~\cite{Hirschi:2019fkz}).

For comparison purposes, we start by providing here the cross section for the hadroproduction of a Higgs boson, computed at LO and NLO QCD in HEFT,
\begin{flalign}
    \sigma^{(\textrm{HEFT},\alpha_s^2\alpha)}_{g g\rightarrow H+X} &= 13.209^{\;+23.4\%\;+2.0\%}_{\;-17.3\%\;-2.0\%} \ \mathrm{pb} \label{eq:LO_HEFT} \\
    \sigma^{(\textrm{HEFT},\alpha_s^2\alpha+\alpha_s^3\alpha)}_{g g\rightarrow H+X} &=
    30.484^{\;+19.8\%\;+1.9\%}_{\;-15.3\%\;-1.9\%} \ \mathrm{pb} \label{eq:NLO_HEFT},
\end{flalign}
where the first set of uncertainties corresponds to the ($\frac{1}{2}$,$1$,$2$) $\mu_R$ scale variation and the second set reports the PDF uncertainty. 
Our final result for the correction from light quarks to the mixed QCD-EW contribution to the inclusive Higgs production cross-section is
\begin{flalign}
    \sigma^{(\alpha_s^2\alpha^2)}_{g g\rightarrow H+X} &= 0.68739\;^{\;+23.4\%\;+2.0\%}_{\;-17.3\%\;-2.0\%} \ \mathrm{pb} \label{eq:LO_EW}\\
    \sigma^{(\alpha_s^2\alpha^2+\alpha_s^3\alpha^2)}_{g g\rightarrow H+X} &=
    1.467(2)^{\;+18.7\%\;+2.0\%}_{\;-14.6\%\;-2.0\%} \ \mathrm{pb}. 
    \label{eq:exactresult}
\end{flalign}
The resulting pure NLO QCD correction of order $\mathcal{O}(\alpha_s^3\alpha^2)$ is $0.780(2)\;\textrm{pb}$ and was obtained from 50K evaluations of the real-emission matrix element in our private implementation of the {\sc\small CoLoRful} NLO subtraction scheme.\\
The gluon initiated cross section with exact EW-boson mass dependence in the virtual contributions, and the real contribution treated in the soft-gluon, the massless and the infinite mass approximation increase the pure gluon-induced HEFT NLO cross section by $5.4\%$\cite{Bonetti:2018ukf}, $5.4\%$ \cite{Anastasiou:2018adr} and $5.2\%$ \cite{Anastasiou:2018adr} respectively, whereas our exact computation, with $\mu_R=\mu_F=\frac{1}{2}M_H$ and $\alpha^{-1}=128.0$ as in ref.~\cite{Bonetti:2018ukf,Anastasiou:2018adr}, yields
$5.1\%$. Our result therefore lies within the original uncertainty assigned to the factorization estimate of $5\pm 1 \%$ given in ref.~\cite{Anastasiou:2002yz} and used in ref.~\cite{Anastasiou:2016cez}. We furthermore notice the small reduction of the scale uncertainty when including NLO corrections. This slow convergence is a known feature of gluon-fusion Higgs production. 

\begin{figure}[ht!]
\centering
\includegraphics[width=\linewidth]{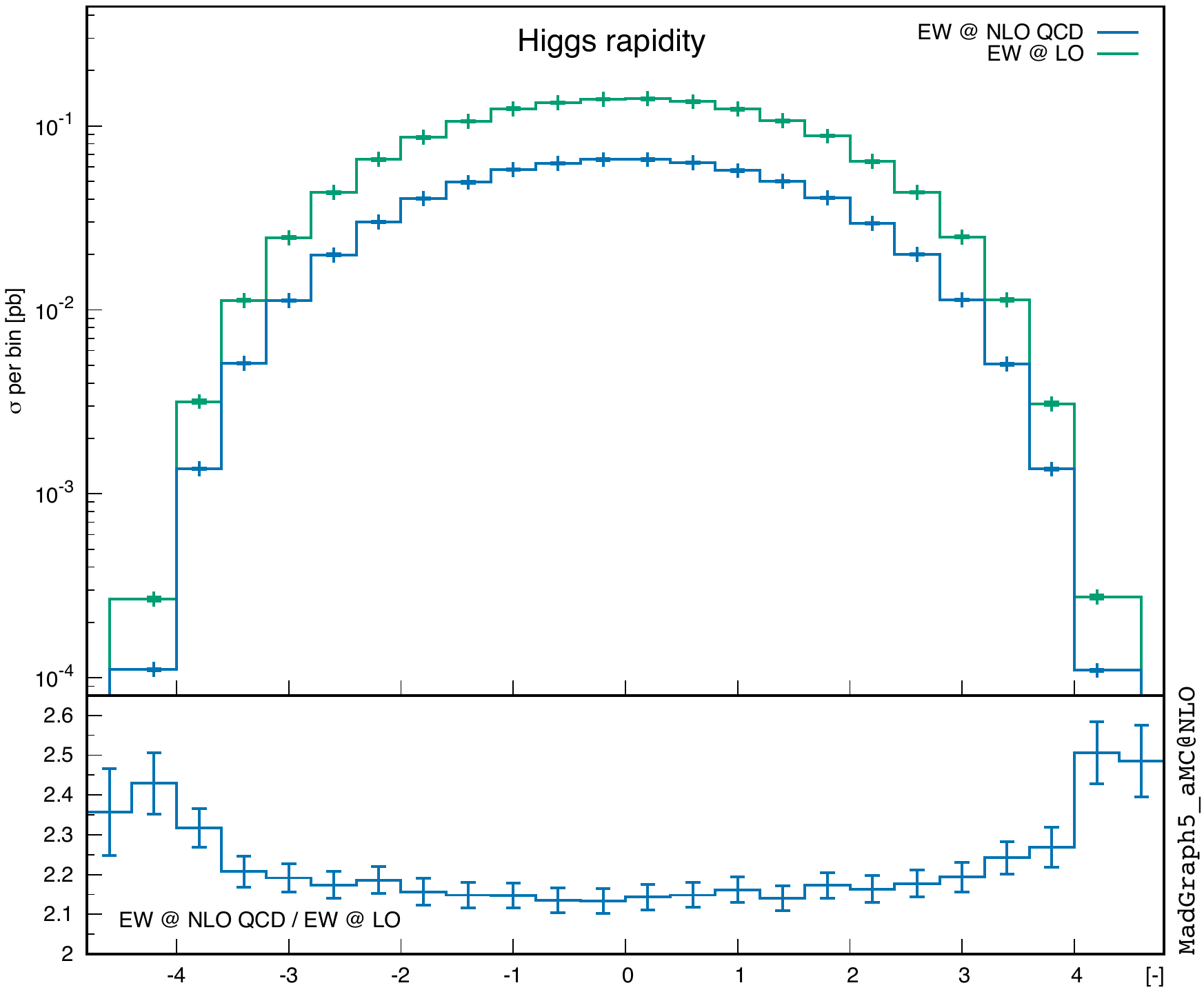}
\caption{\label{fig:rapidity} Differential prediction for the $\mathcal{O}(\alpha_s^3\alpha^2)$ correction to the Higgs rapidity distribution.
}
\end{figure}

\begin{figure}[ht!]
\centering
\includegraphics[width=\linewidth]{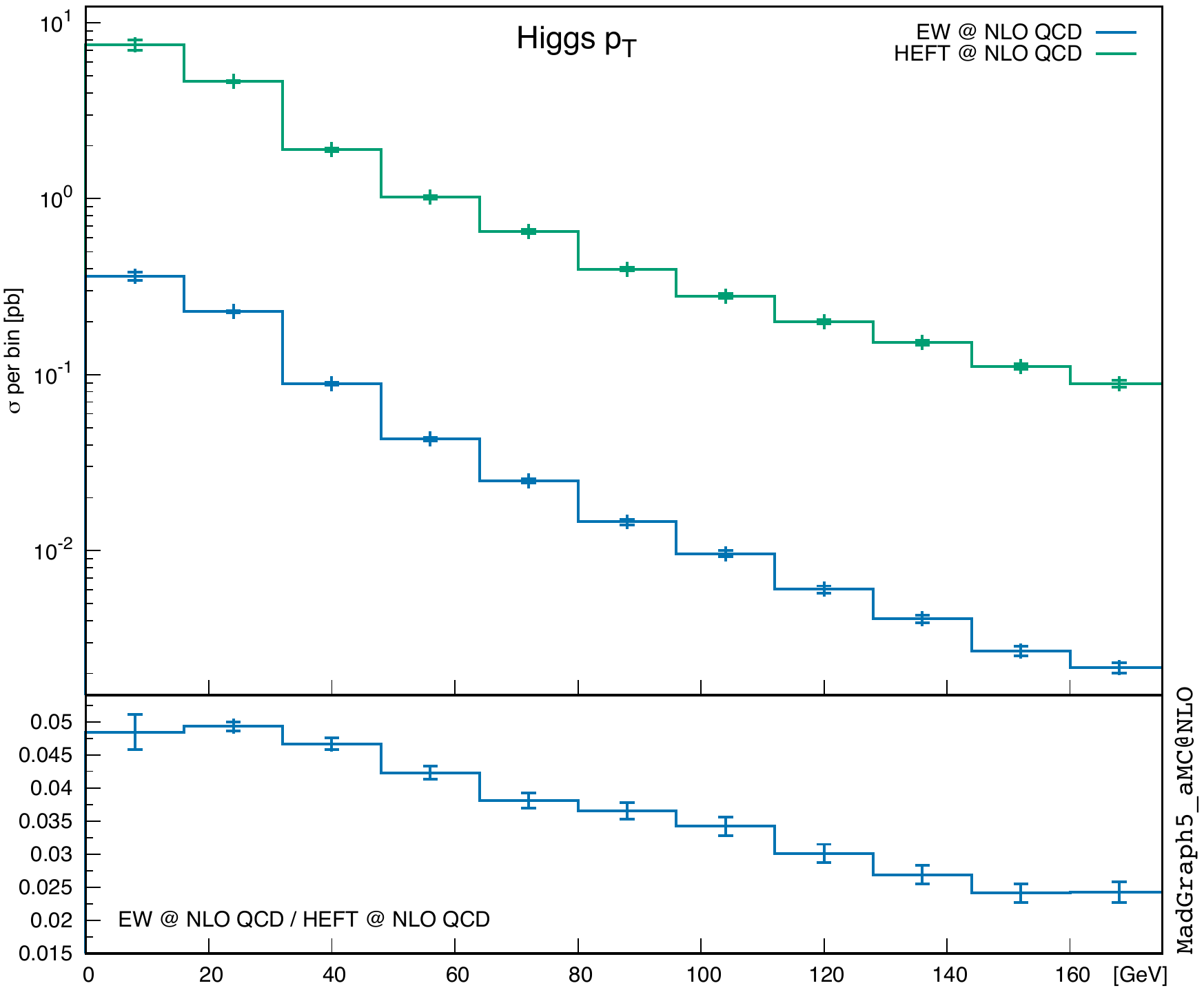}
\caption{\label{fig:higgs_pt} Differential prediction for the $\mathcal{O}(\alpha_s^3\alpha^2)$ EW contribution to the Higgs transverse momentum distribution, compared to its LO HEFT counterpart.
}
\end{figure}

We present the two most relevant differential predictions in figs.~\ref{fig:rapidity} and~\ref{fig:higgs_pt}. First, in fig.~\ref{fig:rapidity}, the NLO-accurate Higgs rapidity distribution, which reveals a flat differential K-factor.
Second, in fig.~\ref{fig:higgs_pt}, the Higgs transverse momentum distribution, which is accurate at leading order, and which we compare to its HEFT counterpart whose spectrum is harder.
We stress that fig.~\ref{fig:higgs_pt} has no direct phenomenological relevance given that quark mass effects (ignored in this work) are poised to affect the shape of the Higgs transverse momentum distributions~\cite{Becker:2020rjp}.

\begin{figure}[ht!]
\centering
\includegraphics[width=\linewidth]{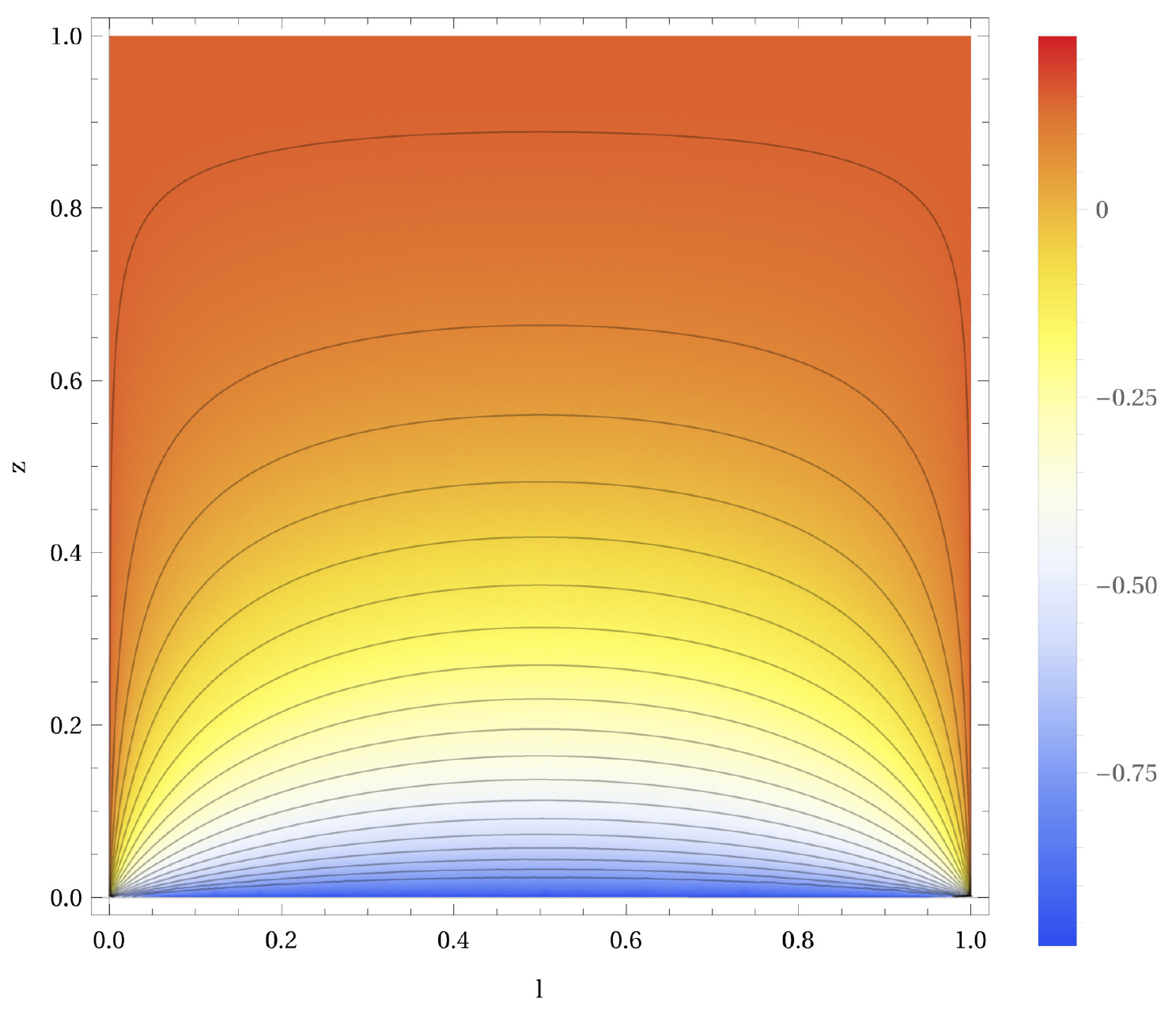}
\caption{\label{fig:HEFTvs2loop} Plot of the quantity {$\frac{\left(
\mathcal{M}_{gg\rightarrow Hg}^{(\alpha_s^3\alpha^2)}/\mathcal{M}_{gg\rightarrow Hg}^{(\alpha_s^3\alpha)}-R^{\text{NLO}}\right)}{R^{\text{NLO}}}$} with ${R^{\text{NLO}}=\sigma^{(\alpha_s^3\alpha^2)}_{g g\rightarrow H+X}/\sigma^{(\textrm{HEFT},\alpha_s^3\alpha)}_{g g\rightarrow H+X}}$ in terms of the rescaled kinematic invariants $z=M_H^2/s$ and $l=t/(M_H^2-s)$, for a sample of $\sim 150$K phase-space points. The lines of constant deviation span the range $[-0.75,0.15]$ in increments of $0.05$.
}
\end{figure}

We also report in fig.~\ref{fig:HEFTvs2loop} on the shape of the kinematic dependence of $\mathcal{M}_{gg\rightarrow Hg}^{(\alpha_s^3\alpha^2)}$ relative to that of $\mathcal{M}_{gg\rightarrow Hg}^{(\alpha_s^3\alpha)}$.
We find that the dependence of this ratio with $z$ is more marked. In particular, we see that the ratio stabilizes rapidly as we approach the production threshold $z\to1$, which amounts to the bulk of the relevant phase-space. This is in line with the observation, that factorization-like approximations provide a good approximation of the total cross-section. For $l\sim0.5$, the limit $z\rightarrow 0$ corresponds to larger values of the Higgs transverse momentum where we see that the two-loop EW matrix element is of smaller magnitude than that of its tree-level HEFT counterpart, as anticipated from the transverse momentum distribution shown in fig.~\ref{fig:higgs_pt}.
\paragraph{The factorization hypothesis}  of the Higgs cross-section is 
\begin{align}
    \sigma_{gg\to H +X}^{(\sum_{i}\alpha_s^{i+2} (\alpha+\alpha^2))} = C^2 \left( \tilde{\sigma}_{gg\to H +X}^{(\alpha_s^2 \alpha)}+  \tilde{\sigma}_{gg\to H +X}^{(\alpha_s^3 \alpha)}+\dots\right)
\end{align}
with 
\begin{align}
    C&=\sum_{i=0} \left(\frac{\alpha_s}{\pi}\right)^i \left( C^{(i)}_{QCD} + \lambda_{EW}  C^{(i)}_{w} \right) 
\end{align}
with $ C^{(0)}_{w}= C^{(0)}_{QCD}=1$, $C^{(1)}_{QCD}=11/4$.
In terms of our quantities from \eqref{eq:LO_HEFT}-\eqref{eq:exactresult}, we have:
\begin{align}
    & \tilde{\sigma}_{gg\to H +X}^{(\alpha_s^2 \alpha)} = {\sigma}_{gg\to H +X}^{(\textrm{HEFT},\alpha_s^2 \alpha)} \\
    &\tilde{\sigma}_{gg\to H +X}^{(\alpha_s^3 \alpha)} = -2 C^{(1)}_{QCD} \sigma^{(\textrm{HEFT},\alpha_s^2\alpha)}_{g g\rightarrow H+X} +\sigma^{(\textrm{HEFT},\alpha_s^3 \alpha)}_{g g\rightarrow H+X} \\
    &\lambda_{EW}=\frac{\sigma^{(\alpha_s^2\alpha^2)}_{g g\rightarrow H+X}}{2 \sigma^{(\textrm{HEFT},\alpha_s^2\alpha)}_{g g\rightarrow H+X}}, \\
    &C^{(1)}_{w}  =C^{(1)}_{QCD} + 
    \left(\frac{\alpha_s}{\pi}\right)^{-1}
    \left( 
    \frac{\sigma^{(\alpha_s^3 \alpha^2)}_{g g\rightarrow H+X}}{ \sigma^{(\alpha_s^2\alpha^2)}_{g g\rightarrow H+X}} 
    - \frac{\sigma^{(\textrm{HEFT},\alpha_s^3 \alpha)}_{g g\rightarrow H+X}}{ \sigma^{(\textrm{HEFT},\alpha_s^2\alpha)}_{g g\rightarrow H+X}} \right).
\end{align}
For $\mu_R=\mu_F$, our exact computation thus yields:
 \begin{align}
     &\lambda_{EW}=0.026, && \\ 
     &C^{(1)}_{w}|_{\mu_R=\frac{1}{2} M_H}=-1.700, && C^{(1)}_{w}|_{\mu_R= M_H}=-2.072 \label{eq:weak_parameters}
 \end{align}
  which is quite different to $C^{(1)}_{w}=7/6$ as estimated in the infinite boson-mass approximation, however still within the uncertainty estimate of ref.~\cite{Anastasiou:2016cez}.
\paragraph{Remaining corrections to our computation}

We identify three main contributions still unaccounted for in our computation of the gluon induced cross-section. We provide here an estimate for each, together with an associated uncertainty:
\begin{itemize}

\item \emph{Heavy-quark (mass) effects} in the QCD amplitudes can be estimated by using
\begin{align}
    K_{\textrm{(N)LO},M_Q}^{\textrm{QCD}}&=\frac{V_{\textrm{FIN},M_Q}^{(\textrm{(N)LO})}}{V_{\textrm{FIN},HEFT}^{(\textrm{(N)LO})}},
    \label{eq:k_factors_qcd_quark_mass}
\end{align}
where the subscript $M_Q$ denotes the QCD background of a heavy quark of mass $M_Q$, ${y_Q=M_Q/v}$ and $V_{\textrm{FIN}}$ as defined in the appendix. The two-loop virtual QCD-amplitude is renormalized in a five-flavour decoupling scheme. We account for heavy quark-mass effects in the QCD-amplitudes by rescaling with $K_{\textrm{(N)LO},M_Q}$ listed in tab.~\ref{tab:k_factors_qcd_quark_mass} and we assign an uncertainty of $\pm 50\%$ of the estimated NLO effect based on the unknown hard real-emission contributions.
\renewcommand*{\arraystretch}{1.6}
\begin{table}
\begin{tabular}{lccc}
 quark & $M_Q [\textrm{GeV}]$ &  $K_{\textrm{LO},M_Q}^{\textrm{QCD}}$   & $K^{\textrm{VIRT,QCD}}_{\textrm{NLO},M_Q}$ \\ \hline 
 charm & 1.3    & -0.010    & -0.018  \\
 bottom & 4.2    &  -0.042   & -0.069 \\
 top & 173    &  \phantom{-}1.032  &  \phantom{-}1.031
\end{tabular}
\caption{K-factors defined as in eq.~\eqref{eq:k_factors_qcd_quark_mass} for different heavy quark masses $M_Q=M_c,M_b,M_t$. \label{tab:k_factors_qcd_quark_mass}}
\end{table}

\item \emph{Top quark effects} in the LO EW-amplitudes were studied in \cite{Degrassi:2004mx,Actis:2008ug} and amount to $-1.8\%$ of the LO cross-section. We assign an uncertainty based on the unknown top quark contribution in the NLO EW amplitudes.

\item \emph{Higher-order QCD corrections} can be accounted under the assumption that the EW form factor receives the same QCD correction as the HEFT operator:
\begin{align}
    \sigma^{(\alpha_s^4+\alpha_s^5)\alpha^2}_{g g\rightarrow H+X} = \sigma^{(\alpha_s^2+\alpha_s^3)\alpha^2}_{g g\rightarrow H+X}\frac{
    \sigma^{(\alpha_s^4+\alpha_s^5)\alpha}_{g g\rightarrow H+X}
    }{
    {\sigma^{(\alpha_s^2+\alpha_s^3)\alpha}_{g g\rightarrow H+X}}
    }
\end{align}
and assign an uncertainty based on testing this EW-QCD factorisation hypothesis on our exact NLO correction. We compute the higher orders in QCD with the program {\sc\small iHixs~2} \cite{Dulat:2018rbf}.
\end{itemize}
When combining the additional contributions and uncertainties above in the setup of tab.~\ref{tableParams}, we arrive at our best estimate for the EW contribution of gluon-initiated Higgs production:
\begin{align}
    \;&\sigma^{(\text{EW},\textrm{best})}_{g g\rightarrow H+X} =
    \sigma^{(\textrm{HEFT},\alpha_s^2\alpha+\alpha_s^3\alpha)}_{g g\rightarrow H+X}
    \times \Big( \nonumber\\
    & \quad \phantom{+\;0} 4.81\;\%\phantom{\pm 0.4\;\%\;}(\text{our computation})\nonumber\\
    & \quad +\;0.15 \pm 0.04\;\%\;(\text{top mass effects in QCD amp.})\nonumber\\
    & \quad -\;0.27 \pm 0.09\;\%\;(\text{bottom quark effects in QCD amp.})\nonumber\\
    & \quad -\;0.07 \pm 0.02\;\%\;(\text{charm quark effects in QCD amp.})\nonumber\\
    & \quad -\;0.04 \pm 0.04\;\%\;(\text{top quark effects in EW})\nonumber\\
    & \quad +\;2.5\phantom{0} \pm 0.4\phantom{0}\;\%\;(\text{QCD higher orders}) \Big )\nonumber \\
    & \quad =\sigma^{(\textrm{HEFT},\alpha_s^2\alpha+\alpha_s^3\alpha)}_{g g\rightarrow H+X}
    \times \Big(\;7.11 \pm 0.6\;\%\ \Big ) \\
    & \quad = 2.17 \pm 0.18 \  \mathrm{pb}.
    \label{eq:bestresult}
\end{align}
A similar computation for $\mu_{R/F}=\frac{1}{2}M_H$ yields {$\sigma^{(\text{EW},\textrm{best})}_{g g\rightarrow H+X} = 2.02 \pm 0.14 \  \mathrm{pb}$}.

\paragraph{Quark-induced and other EW contributions}

The PDF supression from $qq$ production channels render them negligible and we only consider here $qg$-induced channels. We identify the following two categories of quark-induced contributions:
\begin{itemize}
\item \emph{One-loop EW contributions} start at $\mathcal{O}(\alpha_s^2 \alpha^2)$ for the process $q g \rightarrow H q$ and involve one-loop EW triangle and box diagrams interfering with the one-loop QCD Higgs amplitude with exact top and bottom quark mass dependence.
They can be computed exactly using the loop-induced module~\cite{Hirschi:2015iia} of {\sc\small MG5aMC} together with the loop-ready EW {\sc\small UFO} model of ref.~\cite{Frederix:2018nkq}. We however use here the result of tab.~3 of ref.~\cite{Hirschi:2019fkz} which includes \emph{all} one-loop EW contributions in the $gg$ and $qg$ channels (incl. Higgs-strahlung). We assign a theoretical uncertainty to this contribution based on the $38\%$ $\mu_R$ scale variation of its $qg$ component of order $\mathcal{O}(\alpha_s^2\alpha^2)$ and we obtain:
\begin{align}
    \;&\sigma^{(\text{EW})}_{\textrm{1-loop}} = 
    \begin{cases}
     0.025 \pm 0.052 \  \mathrm{pb} \ ; \ \mu_{R/F}=M_H \\
     0.031 \pm 0.062 \  \mathrm{pb} \ ; \ \mu_{R/F}=\frac{M_H}{2}
     \end{cases}.\nonumber
\end{align}
The large uncertainty assigned here reflects the accidental cancellation found among all contributions considered in ref.~\cite{Hirschi:2019fkz}.

\item \emph{Quark corrections to EW form-factors} correspond to $qg$ contributions stemming from the QCD evolution of the initial-state gluon entering the EW two-loop form factor. We estimate this contribution by rescaling our LO EW result by the $qg$ K-factor of the HEFT contribution and assign an uncertainty of 10\% based on the relative difference of the $\mu_F$ dependence of the NLO EW and NLO HEFT gluon-induced contribution, and a 30\% uncertainty based on the unknown hard real emission contribution. To estimate the higher order QCD-corrections we perform the same rescaling and assign an uncertainty of 50\% based on the fact, that we do not know if they are captured well by a factorization ansatz. We obtain:
\begin{align}
    \;&\sigma^{(\text{EW})}_{q g\rightarrow H+X} = 
    \begin{cases}
     -0.10 \pm 0.05 \  \mathrm{pb} \ ; \ \mu_{R/F}=M_H \\
     \phantom{-}0.12 \pm 0.05 \  \mathrm{pb} \ ; \ \mu_{R/F}=\frac{M_H}{2}
     \end{cases},\nonumber
\end{align}
where we used the program {\sc\small iHixs~2} to compute the QCD corrections beyond NLO.
Notice that, as expected, including $\sigma^{(\text{EW})}_{q g\rightarrow H+X}$ helps reducing the difference in the quantity $\sigma^{(\text{EW},\textrm{best})}_{g g\rightarrow H+X}$ when computed with $\mu_F=M_H$ and $\mu_F=\frac{M_H}{2}$.
\item \emph{Higher order EW} contributions are estimated by re-weighting our cross-section with
\begin{align}
    K_{\textrm{(N)LO}}^{\textrm{pure-EW}}&=\frac{V_{\textrm{FIN},\textrm{pure-EW}}^{(\textrm{(N)LO})}}{V_{\textrm{FIN},\textrm{mixed-QCD-EW}}^{(\textrm{(N)LO})}},
\end{align}
where we compute $V_{\textrm{FIN},EW}^{(\textrm{(N)LO})}$ by squaring (interfering)  the relevant 2- and 3-loop EW virtual amplitudes and assign a 50\% uncertainty on the NLO-result based on the unknown hard-real radiation:
\begin{align}
    \;&\sigma^{(\text{squared EW})}_{g g\rightarrow H+X} = 
    \begin{cases}
      0.018 \pm 0.005 \  \mathrm{pb} \ ; \ \mu_{R/F}=M_H \\
     0.020 \pm 0.005 \  \mathrm{pb} \ ; \ \mu_{R/F}=\frac{M_H}{2}
     \end{cases}.\nonumber
\end{align}

\end{itemize}
Together with the gluon-induced contributions of eq.~\eqref{eq:bestresult}, we finally construct our best estimate for the overall contributions of EW origin to Higgs hadro-production with the parameters given in tab.~\ref{tableParams}:
\begin{align}
     \;&\sigma^{(\text{EW},\textrm{best})}_{p p\rightarrow H+X} \nonumber \\
     &\ =
     \begin{cases}
        (6.91 \pm 0.9 \% )\times\sigma^{(\textrm{HEFT},\alpha_s^2\alpha+\alpha_s^3\alpha)}_{g g\rightarrow H+X} \ ; \ \mu_{R/F}=M_H \\
        (6.43 \pm 0.8 \% )\times\sigma^{(\textrm{HEFT},\alpha_s^2\alpha+\alpha_s^3\alpha)}_{g g\rightarrow H+X} \ ; \ \mu_{R/F}=\frac{M_H}{2}
     \end{cases} \nonumber \\
     &\ =
         \begin{cases}
            \phantom{(}2.11 \pm 0.28 \ (\textrm{theory}) \ \mathrm{pb} \ ; \ \mu_{R/F}=M_H \\
            \phantom{(}2.19 \pm 0.26 \ (\textrm{theory}) \ \mathrm{pb} \ ; \ \mu_{R/F}=\frac{M_H}{2}
     \end{cases}.
\end{align}

\section{Conclusion}
We evaluated the NLO QCD correction to mixed EW-QCD light quark contribution to Higgs production via gluon fusion.
Unlike previous computations of this quantity, we retained the exact dependence on the weak boson masses.
The two-loop real-emission amplitudes are computed by solving differential equations for the relevant scalar integrals at runtime in terms of one-dimensional generalised power series. We implemented the resulting matrix-elements in a flexible manner by encoding them as form factors of a {\sc\small UFO} model, which we make publicly available as a {\sc\small MadGraph5\_aMC@NLO} plugin.
We performed the phase-space integration numerically in two separate implementations of the {\sc\small FKS} and {\sc\small CoLoRFuL} subtraction scheme using an offline parallelisation model in order to accommodate the evaluation speed of our matrix elements.

We presented the distribution of the Higgs rapidity, which shows a flat differential K-factor, and of the Higgs transverse momentum, whose spectrum is softer than its HEFT counterpart when ignoring quark mass effects.

When related to the NLO accurate HEFT cross-section, we find the EW contribution from light quarks to be $5.1\%$ for $\mu_R=\mu_F=\frac{1}{2}M_H$, which is very close to the result of $5.2\%$ obtained in the infinite weak boson mass limit~\cite{Anastasiou:2018adr,Anastasiou:2002yz,Anastasiou:2016cez}.
Our result therefore allows to reduce the uncertainty of $\pm 1\%$ assigned thus far to the EW gluon-induced contribution in order to reflect the absence of an exact weak boson mass treatment at NLO.
We also conclude that violations to the factorisation of EW and QCD contributions are modest. This gives further confidence in the rescaling of our cross-section with higher-order QCD corrections computed in HEFT which, together with the estimate of other partially unknown effects, yields our best estimate of $6.91 \pm 0.93 \% $ (relative to the gluon-induced NLO HEFT) for the overall contribution of EW origin to Higgs production at the LHC. The LO and NLO quark-induced EW contribution now stands as an important source of uncertainty, and they can be computed in the future using the same methods as those presented in this letter.

\section{Acknowledgements}
This project has received funding from the European Research Council (ERC) under grant agreement No 694712 (PertQCD) and the Swiss  National  Science  Foundation (SNF) under contract agreement No 177632. We thank S. Abreu for useful discussions about cut Feynman integrals, N. Deutschmann for his collaboration on the development of the implementation of the {\sc\small CoLoRful} subtraction and many useful discussions, A. Ochirov for insightful discussions about helicity amplitudes, A. Lazopoulos for useful discussion regarding iHixs, and
C. Anastasiou for his valuable feedback on the draft of this letter.

\appendix
\section{Benchmark numerical matrix element evaluations}
\label{sec:benchmar_results}

In order to facilitate the reproduction of our work, we report here benchmark numerical results of our matrix elements, summed/averaged over helicity and color configurations, evaluated for specific kinematic configurations with $\alpha_S(M_Z)=0.118$ and all other SM parameters as in table~\ref{tableParams}.
For the virtual matrix element, we only report the finite part as defined in eq.~(B.2) of ref.~\cite{Frederix:2009yq} or eq.~(A.1) of ref.~\cite{Hirschi:2011pa}.
The $gg\rightarrow H$ matrix elements are:
\begin{eqnarray}
V_{\textrm{FIN}}^{(\textrm{LO})}&&=\overline{\sum_{h,c}} 2\Re\left[
\mathcal{A}^{(2-\textrm{loop})}_{gg\rightarrow H}\mathcal{A}^{(\textrm{tree})\star}_{\textrm{HEFT}}
\right]\nonumber\\
&&=\texttt{5.1508192663885}\cdot\texttt{10}^{\texttt{-4}} \left[\texttt{GeV}^\texttt{2}\right] \\
V_{\textrm{FIN}}^{(\textrm{NLO})}&&=\overline{\sum_{h,c}} 2\Re\left[
\mathcal{A}^{(3-\textrm{loop})}_{gg\rightarrow H}\mathcal{A}^{(\textrm{tree})\star}_{\textrm{HEFT}}
+\mathcal{A}^{(2-\textrm{loop})}_{gg\rightarrow H}\mathcal{A}^{(1-\textrm{loop})\star}_{\textrm{HEFT}}
\right]_\textrm{FIN}\nonumber
\\&&=\texttt{3.6824078313996}\cdot\texttt{10}^{\texttt{-4}} \left[\texttt{GeV}^\texttt{2}\right] .
\end{eqnarray}
We also provide the relation of $V_{\textrm{FIN}}^{(\textrm{NLO})}$ above with the matrix element $M_{\textrm{FIN},I_1}$ obtained from the finite amplitudes computed with the methods of refs.~\cite{Bonetti:2016brm,Bonetti:2017ovy}:
\begin{eqnarray}
    V_{\textrm{FIN}}^{(\textrm{NLO})}= M_{\textrm{FIN},I_1}+C_A \pi^2 V_{\textrm{FIN}}^{(\textrm{LO})},
\end{eqnarray}
where $C_A=3$ denotes the number of colors.\\
Finally, we give the 2-loop $gg\rightarrow Hg$ matrix elements evaluated at two different benchmark kinematic points specified with the two Mandelstam invariants $s=(p_{g_1}+p_{g_2})^2$ and $t=(p_{g_2}-p_{H})^2$:
\begin{eqnarray}
&&\overline{\sum_{h,c}} 2\Re\left[\mathcal{A}^{(2-\textrm{loop})}_{gg\rightarrow Hg}\mathcal{A}^{(\textrm{tree})\star}_{\textrm{HEFT}}\right](s=5M_H^2,t=-3M_H^2)\nonumber\\
&&=\texttt{9.0303320385123}\cdot\texttt{10}^{\texttt{-6}} \left[\texttt{GeV}^\texttt{0}\right] \\
&&\overline{\sum_{h,c}} 2\Re\left[\mathcal{A}^{(2-\textrm{loop})}_{gg\rightarrow Hg}\mathcal{A}^{(\textrm{tree})\star}_{\textrm{HEFT}}\right](s=5M_H^2,t=-2M_H^2)\nonumber\\
&&=\texttt{5.8988801633472}\cdot\texttt{10}^{\texttt{-6}} \left[\texttt{GeV}^\texttt{0}\right]
\end{eqnarray}

\bibliographystyle{apsrev4-1}
\bibliography{refs}

\end{document}